\documentclass[preprint,12pt]{elsarticle}

\usepackage{svg}  
\usepackage{amssymb}
\usepackage{amsmath}

\usepackage{multirow} 

\usepackage{color}

\usepackage{soul}

\begin{document}
\begin{frontmatter}



\title{One predator and two prey: Coexistence of pumas, guanacos and sheep in Patagonia}

\author[label1]{Jhordan Silveira de Borba \corref{cor1}}
\ead{sbjhordan@gmail.com}
 \cortext[cor1]{Corresponding author.}
\author[label1]{Sebastian Gonçalves}
\ead{sgonc@if.ufrgs.br}

\affiliation[label1]{organization={Instituto de Física, Universidade Federal do Rio Grande do Sul},
            addressline={Av. Bento Gonçalves 9500}, 
            city={Porto Alegre},
            postcode={91501-970}, 
            state={RS},
            country={Brazil}}

\begin{abstract}
The ecosystem considered in this study is the outcome of a lengthy sequence of historical and ecological events. Patagonia's indigenous fauna comprises survivors of five significant extinction events, with the notable presence of the puma and the guanaco, two of the largest native mammals. In addition to these, European immigrants introduced sheep into the ecosystem. Together, these three species form a straightforward trophic network, featuring one predator and two prey species, all competing within the Patagonian steppe.
For ranchers, guanacos and pumas are frequently perceived as threats to their economic interests, making them targets for ongoing removal through hunting. In recent decades, the field of biology, particularly ecology, has witnessed a substantial increase in the development of equation-based models. Scientists are interested in the ability to systematize hypotheses and gain insights into the behavior of complex biological systems, such as the one presented in this study.
However, the nonlinear nature and the large number of parameters of models, represent a challenge when one wants to explore the parameter space.  To overcome this and, at the same time, improve the understanding of the Patagonia ecosystem, we start by building an equation-based model 
based on previous contributions, and we reduce it to the essential minimum set of parameters. Then, we introduce two tools, a generalization of ternary graphs and a perceptron based ML, to help understand the response of the system equation to  the key parameters.
The perceptron tool allows us to visualize/interpret the influence of each parameter on the survival or extinction of each species.  Through the generalization of the ternary graph, it was possible to conveniently visualize how the system responds to different combinations/variations of the five parameters of the reduced system equation in a single graphical representation.
\end{abstract}


\end{frontmatter}

\section{Introduction}

Mathematical modeling of ecological interactions is an essential tool for comprehending and, in the best possible way, predicting the behavior of systems of coexisting species. The interrelationship of factors as diverse as climate, access to resources, predators, and human activity makes it necessary to develop mathematical models that allow the prediction of the effect of each of the species involved, indicating possible scenarios of coexistence or extinction~\cite{laguna3}. The literature is full of examples of different models, predator-prey models~\cite{reflaguna1,reflaguna2,reflaguna3}, intra- and inter-specific competition~\cite{reflaguna4,reflaguna5,reflaguna6}, and more~\cite{reflaguna7,reflaguna8,reflaguna9,reflaguna10,reflaguna11}. However there is a difficulty in studying each of these models exhaustively. It is commonplace to restrict the discussion either to a few specific cases or to narrow down the system's reactions when exposed to variations in a handful of parameters.

This contribution focuses on a comprehensive exploration of the dynamics of a simple trophic web. Specifically, we analyze a system consisting of a single predator and two competing prey species in the Patagonian steppe. The study examines two native species—puma (\textit{Puma concolor}, a carnivore) and guanaco (\textit{Lama guanicoe}, a camelid)—alongside sheep (\textit{Ovis aries}), an introduced competitor of the native herbivore and an additional prey for pumas. The mathematical model developed in this work is inspired by the history of environmental degradation in Patagonia, which stems from a long sequence of ecological and historical events briefly outlined below.

The survivors of the five main extinctions constitute the indigenous mammalian fauna of Patagonia. The last major event occurred during the quaternary glaciations, when both climate change and the arrival of humans occurred for the first time in the continent's evolutionary history~\cite{reflaguna13}. Pumas and guanacos, currently the two largest mammals in Patagonia, have coexisted with humans for at least 13,000 years, with no signs of their ranges shrinking until the twentieth century~\cite{reflaguna14,reflaguna15,reflaguna16}.

Historical records indicate a great abundance of guanacos coexisting sustainably with Tehuelche hunters until shortly before the arrival of European immigrants~\cite{reflaguna17,reflaguna18}. The Tehuelche people built their ecological niche by learning to follow massive groups of guanacos, up to 300,000 individuals, during their seasonal migrations, maintaining a nomadic system that persisted sustainably for at least 6,000 years~\cite{reflaguna17}. However, during the "Wingka Malón" in Argentina (1880–1890), the national army nearly exterminated or completely expelled the highly mobile indigenous hunters from their ancestral territories~\cite{reflaguna19}.
The division of the Patagonian steppe into large and continuous extensions of land gave rise to private ranches, organized by a gigantic network of fences, where sheep farming was the economic activity of 95\% of these private farms~\cite{reflaguna20}. The introduction of sheep has then significantly altered the ecological interactions of Patagonian flora and fauna~\cite{reflaguna21}. Ranchers, descendants of Europeans, created a new niche in which the puma, as a predator of sheep, and the guanaco, as a competitor for forage, became enemies of their economic interests and consequently targeted for eradication~\cite{reflaguna20}.

Ranches work to maximize the use of biomass by sheep flocks~\cite{lagunasavory} often focused on short-term gain~\cite{lagunaottichilo}. Previous studies have shown that the abundance of wild herbivores is inversely correlated with the density of livestock due to direct and indirect competition~\cite{lagunabaldi,lagunapedrama}. There is evidence of competition between sheep and guanacos~\cite{reflaguna22,reflaguna20}, mainly for food and water. Of a diet of 80 plant species, they share 76~\cite{reflaguna23}. Under natural conditions, the guanaco is a superior competitor to sheep, able to displace sheep from water sources (including artificial ones).

As fields deteriorate due to overgrazing and desertification, the guanaco increases its competitive superiority over sheep, since it is superbly adapted to situations of environmental harshness, especially water scarcity (e.g., it can drink sea water in case of extreme necessity). Therefore, the density of guanacos naturally increases when the productivity of the fields decreases. However, this natural process is usually offset by an increased hunting pressure on guanacos as environmental conditions worsen, as ranchers seek to maximize scarce resources for production. Droughts exacerbate these socio-environmental crises, as the lack of rain undermines the carrying capacity for both wildlife and livestock. Pumas, as predators of sheep, are also subjected to constant culling in efforts to reduce production costs~\cite{reflaguna20}.  The puma naturally preyed on guanacos~\cite{reflaguna24}, but since the introduction of sheep, it has shifted almost entirely to hunting them. Sheep require significantly lower exploration costs ---that is, the energy expended to locate, pursue, and capture prey--- compared to the high energy demands of hunting the fast and elusive guanaco, which co-evolved with the puma~\cite{reflaguna25}.

Given that inadequate livestock management ---typically involving more animals than the land's carrying capacity--- is the norm, habitat degradation, loss of nutritionally valuable species, and wildlife extinction are common outcomes~\cite{lagunagolluscio,lagunafunlendorf}. Understanding how these anthropogenic factors influence decision-making in sheep production management and its coexistence with wildlife is crucial~\cite{laguna4}.

In this contribution, we study a mathematical model that represents a simplified version of the puma-guanaco-sheep ecosystem. Our goal is to provide a theoretical framework for formalizing various field observations and conceptual models. At this stage, we deliberately keep the model simple to enable an exhaustive exploration of the system's behavior. The next section introduces the equation-based model, followed by an analysis of the main results obtained using different tools, including a variation of the ternary graph and artificial neural networks. We believe that these tools can enhance the understanding of other ecological models. The final section discusses the results and outlines implications for future research directions.

\section{Analysis}
\subsection{Meta-population model}
\label{MBE}
Inspired by the ecosystem of Argentinian Patagonia, which features coexisting populations of sheep, guanacos, and pumas, previous studies~\cite {laguna1}
have proposed the following system of equations to model their interactions:

\begin{equation}
    \begin{split}
        \dot{x}_{1}&=c_{1}x_{1}\left(1-D-x_{1}\right)-e_{1}x_{1}-\mu_{1} x_{1}y\\
        \dot{x}_{2}&=c_{2}x_{2}\left(1-D-x_{1}-x_{2}\right)-e_{2}x_{2}-\mu_{2}x_{2}y-c_{1}x_{1}x_{2}\\
        \dot{y}&=c_{y}y\left(x_{1}+x_{2}-x_{1}x_{2}-y\right)-e_{y}y.
    \end{split}
    \label{original_completo}
\end{equation}
The system of Eqs.~\ref{original_completo} represents a spatially implicit model 
inspired by meta-populations distributed on a grid. Analogous to Levins’ model, the variables $x_1$ and $x_2$ represent the proportion of space occupied by the prey species, guanacos and sheep, respectively, while $y$ represents the predator, the puma.
The system has nine parameters, whose interpretation we describe below:
\begin{itemize}
        \item $D$ represents the fraction of the destruction of the ecosystem
    \item $c_j$ is the rate of colonization of the species, $j \in \{1,2,y\}$
    \item $e_j$ local extinction rate of species, $j \in \{1,2,y\}$
    \item $\mu_j$ rate of predation (predator on preys), $j \in \{1,2\}$
\end{itemize}

A recent paper for an analogous system with one predator and two prey in a hierarchical competition was originally published with a typo~\footnote{Private communication with authors.}: the absence of the term $c_y y^2$ in the predator equation~\cite{laguna2}.  This typo inspired us to simplify the system by removing it. This term represents the limit of growth of predators competing for resources, which in practice never happens because the growth of the puma population is ultimately limited by human action. With this change, the system is:
\begin{equation}
    \begin{split}
        \dot{x}_{1}	& =c_{1}x_{1}\left(1-D-x_{1}\right)-e_{1}x_{1}-\mu_1 x_{1}y \\
\dot{x}_{2}	& =c_{2}x_{2}\left(1-D-x_{1}-x_{2}\right)-e_{2}x_{2}-\mu_{2}x_{2}y-c_{1}x_{1}x_{2} \\
\dot{y}	& =c_{y}y\left(x_{1}+x_{2}-x_{1}x_{2}\right)-e_{y}y.
    \end{split}
    \label{laguna2_completo}
\end{equation}

We propose a final simplification by eliminating the term $c_y x_1 x_2 y$ from the predator equation. The original metapopulation model depicts predator colonization of new cells only when they are already occupied by sheep or guanacos, similar to a model of cellular automata.  However, since prey can simultaneously occupy the same cell, it is necessary to discount this value to avoid double counting. Without this analogy, this term loses its primary utility and becomes a suitable candidate for removal to simplify our system. This adjustment results in the predator population decrease being linearly proportional only to its own population. Therefore, we obtain the following revised model:
\begin{equation}
    \begin{split}
        \dot{x}_{1} & =	c_{1}x_{1}\left(1-D-x_{1}\right)-e_{1}x_{1}-\mu_{1}x_{1}y \\
        \dot{x}_{2} & =	c_{2}x_{2}\left(1-D-x_{1}-x_{2}\right)-e_{2}x_{2}-\mu_{2}x_{2}y-c_{1}x_{1}x_{2} \\
        \dot{y}    & =	c_{y}y\left(x_{1}+x_{2}\right)-e_{y}y.
    \end{split}
    \label{modelo completo}
\end{equation}
Rearranging the terms of Eq.~\ref{modelo completo}, where $H=1-D$ represents the fraction of available territory for species colonization, we obtain a slightly simple set of equations:
\begin{equation}
    \begin{split}
        \dot{x}_{1} & =	a_{1}x_{1}-c_{1}x_{1}^{2}-\mu_{1}x_{1}y \\
\dot{x}_{2} & =	a_{2}x_{2}-c_{2}x_{2}^{2}-\mu_{2}x_{2}y-\mu_{12}x_{1}x_{2} \\
\dot{y} & =	a_{y}\left(x_{1}+x_{2}\right)y-e_{y}y
    \end{split}
    \label{nosso_completo}
\end{equation}

The new parameters in the system Eqs.~\ref{nosso_completo} are expressed in terms of the original ones, as shown in Table~\ref{tabela} below.

\begin{table}[ht]
\caption{Parameters of Eqs.~\ref{nosso_completo} in terms of parameters of Eqs.~\ref{modelo completo}.}

\label{tabela}
\centering
\begin{tabular}{|l|l|}
\hline
Eqs.~\ref{nosso_completo} & Eqs.~\ref{modelo completo} \\
\hline
$a_1$ & $c_{1}H-e_{1}$ \\
$a_2$ & $c_2H-e_2$ \\
$a_y$ & $c_y$ \\ 
$\mu_{12}$ & $c_1+c_2$ \\ 
\hline
\end{tabular}
\end{table}

These parameters maintain an analogous interpretation to their predecessors:
\begin{itemize}
    \item $a_j$ represents the reproduction rate of species $j$
    \item $\mu_{12}$ indicates the hierarchical competition rate between herbivore species
\end{itemize}
We can further simplify the system to a dimensionless set of equations by retaining only the essential independent parameters. Applying four scaling transformations results in an equivalent system with just five relevant parameters. The first transformation consists of dividing the system by $a_2$, noting that the transformation $\widehat{t}=a_2t$ represents a change in the time scale; subsequently, we transform each independent variable by defining $\widehat{x}_{j}=\frac{a_{y}}{a_{2}}x_{j}$ and $\widehat{y}=\frac{\mu_{2}}{a_{2}}y$.
The resulting equation, Eqs.~\ref{quase_final}, and the new parameters are read in terms of the relationships between the previous ones, as can be seen in Table~\ref{tabela2},
\begin{equation}
    \begin{split}
        \frac{d \widehat{x}_{1}}{d\widehat{t}} & =	a\widehat{x}_{1}-\frac{\widehat{x}_{1}^{2}}{\kappa_{1}}-\mu\widehat{x}_{1}y \\
\frac{d \widehat{x}_{2}}{d\widehat{t}} & =	\widehat{x}_{2}-\frac{\widehat{x}_{2}^{2}}{\kappa_{2}}-\widehat{x}_{2}\widehat{y}-p\widehat{x}_{1}\widehat{x}_{2} \\
\frac{d \widehat{y}}{d\widehat{t}} & =	\left(\widehat{x}_{1}+\widehat{x}_{2}\right)\widehat{y}-e\widehat{y}.
    \end{split}
    \label{quase_final}
\end{equation}
\begin{table}[ht]
\caption{Essential independent, dimensionless parameters of Eqs.~\ref{eq_final} in terms of the original ones of Eqs.~\ref{nosso_completo}.}
\label{tabela2}
\centering
\begin{tabular}{|l|l|}
\hline
Eqs.~\ref{eq_final} & Eqs.~\ref{nosso_completo} \\
\hline
$a$ & $a_1/a_2$ \\ 
$\mu$ & $\mu_1/\mu_2$\\ 
$e$ & $e_{y}/a_{2}$ \\ 
$\kappa_1$ &  $a_y/c_1$\\ 
$\kappa_2$ & $a_y/c_2$ \\ 
\hline
\end{tabular}
\end{table}
where $p$ depends on the parameter, so it is not independent:
\begin{equation}
p=\frac{\mu_{12}}{a_{y}}=\frac{c_{2}}{a_{y}}+\frac{c_{1}}{a_{y}}=\frac{1}{\kappa_{1}}+\frac{1}{\kappa_{2}}
\end{equation}

So after the transformations, we take the original nine-parameter system to an equivalent one with five essential parameters. For ease of reading, we drop the {\em hat} from the notation ($\hat{x} \equiv x$), so we have the final set of equations: 
\begin{equation}
    \begin{split}
        \dot{x}_{1} & =	ax_{1}-\frac{x_{1}^{2}}{\kappa_{1}}-\mu x_{1}y \\
        \dot{x}_{2} & =	x_{2}-\frac{x_{2}^{2}}{\kappa_{2}}-x_{2}y-px_{1}x_{2} \\
        \dot{y}     & =	\left(x_{1}+x_{2}\right)y-ey
    \end{split}
    \label{eq_final}
\end{equation}

\subsection{Equilibria and state-space representation}
\label{equiibrio}
In this section, we determine the equilibrium points of the system equations and analyze their stability. This approach allows us to ascertain the survival outcomes of the animal populations once the system reaches an equilibrium state, eliminating the need for numerical solutions to understand the system's evolution.

The equilibrium of Eqs.~\ref{eq_final} corresponds to the three derivatives being equal to zero, {\em i.e} 
$\boldsymbol{\dot{x}}=\left(0,0,0\right)$ giving rise to three algebraic equations that must be zero simultaneously. Then the system can be factorized as:
\begin{equation}
    \begin{split}
        0 & =	x_1 \left(a - \frac{x_1}{\kappa_1} - \mu y\right) \\
        0 & = x_2 \left( 1 - \frac{x_2}{\kappa_2} - y - px_1 \right) \\
        0 & = y	\left( x_1+x_2 - e \right) 
    \end{split}
    \label{equilibria}
\end{equation}
As each of the three species can either be present (not equal to 0) or absent (equal to 0), we end up with eight combinations of conditions that satisfy these equations, representing different types of equilibrium. Some of these equilibria are trivial and hold no ecological interest, such as ${\boldsymbol{x}}_{0}=\left(0,0,0\right)$, which corresponds to the absence of all species.

Next, we have three {\em single-species} equilibria, where the other two variables are set to zero. For example, ${\boldsymbol{x}}_{1}=\left(x_{1}^{*},0,0\right)$  corresponds to a {\em guanaco-alonesolution } with $x_1^* = a \kappa_1$. A similar solution exists for {\em sheep-alone} ($x_2^* = \kappa_2$). However, the {\em puma-alone} equilibrium  ${\boldsymbol{x_3^*}}=\left(0,0,y^*\right)$ is not feasible, as pumas require prey to survive~\footnote{The mathematical condition in this case is $y^* e  = 0$, implying $e=0$, which would be an unrealistic situation of pumas not dying even without food.}.

The following three nontrivial equilibria have ecological interest and correspond to the coexistence of two of the three species. First, we consider the coexistence of the prey, without the predator, ${\boldsymbol{x}}_{4}=\left(x_{1}^{*},x_{2}^{*},0\right)$, so we have:
\begin{equation}
    \begin{split}
        0 & = a-\frac{x_{1}^*}{\kappa_{1}} \\
        0 & = 1-\frac{x_{2}^*}{\kappa_{2}}-p x_{1}^*\;, \\
        \label{presas}
    \end{split}
\end{equation}
from where we arrive at ${\boldsymbol{x}}_{4}=\left( a \kappa_1,\kappa_{2}(1 -p a\kappa_{1}),0\right)$. In this equilibrium, sheep coexists with guanaco, but while the latter maintains the same occupancy that it has alone, the occupancy of the sheep is reduced by a fraction $p a\kappa_1$, due to the pressure of guanaco. Similarly, we have two solutions of coexistence of one of the prey and the predator, which results in the following equilibrium points:
\begin{itemize}
    \item ${\boldsymbol{x}}_{5}=\left(x_{1}^{*},0,y^*\right)=\left(e,0,\frac{a}{\mu}-\frac{e}{\mu \kappa_{1}}\right)$
    \item ${\boldsymbol{x}}_{6}=\left(0,x_{2}^{*},y^*\right)=\left(0,e,1-\frac{e}{\kappa_{2}}\right)\; , $
\end{itemize}
where the first corresponds to the coexistence of puma and guanaco, and the second to the coexistence of puma and sheep. In terms of the original variables and parameters, the guanacos and pumas equilibrium is:

$$x_1^* = \frac{e_y}{a_2}\;,\;\; y^{*}=\frac{\mu_{2}}{\mu_{1}}(\frac{a_{1}}{a_{2}}-\frac{e_{y}}{a_{2}}\frac{c_{1}}{a_{y}})$$
then
$$x_1^* = \frac{e_y}{c_2H - e_2}$$
$$y^{*}=\frac{\mu_{2}}{\mu_{1}}\frac{e_{1}}{\left(c_{2}H-e_{2}\right)}\left[\frac{c_{1}}{e_{1}}\left(H-\frac{e_{y}}{c_{y}}\right)-1\right]$$

We can see that there is a relation between the parameters and what is expected. For example, looking at the equilibrium point for the guanaco ($x_1^*$), it is higher the higher the predator extinction rate ($e_y$); similarly, the population of guanaco in equilibrium decreases as we increase the reproduction rate of sheep ($c_2$). Similarly, it increases as the rate of extinction of sheep increases ($e_2$). A more unexpected behavior concerns the non-destroyed fraction of the system ($H$), since given a set of parameters that satisfy the conditions for this equilibrium point to be reached, it seems to contribute inversely to the final guanaco population at equilibrium, i.e., the larger the available fraction, the smaller the final population.

Looking at the predator, we notice that both parameters that decrease the survival of the prey, the predation rate ($\mu_1$) and the extinction rate ($e_1$), also decrease the population of pumas in equilibrium ($y^*$). On the other hand, the reproduction rate of the prey ($c_1$) contributes to an increase in the final population of pumas. There is also a relationship between the extinction rate ($e_y$) and the reproduction rate ($c_y$) of the predator, where, as expected, the extinction rate decreases the final population, while the reproduction rate increases. Regarding the sheep, the predation rate ($\mu_2$) and the extinction rate ($e_2$) promote an increase in the puma population, while the reproduction rate ($c_2$) hinders it.  The reaction of the puma population to the number of patches destroyed ($H$) will depend on the relationship between the other parameters.

And about sheep and pumas in equilibrium:

\begin{equation}
x_2^* = \frac{e_y}{a_2} \;,\;\;y^*=1-\frac{e_{y}}{a_{2}}\frac{c_{2}}{a_{y}}
\end{equation}
then
$$x_2^* = \frac{e_y}{c_2H - e_2}$$
\begin{equation}
    y^{*}=1-\frac{1}{\left(H-\frac{e_{2}}{c_{2}}\right)}\frac{e_{y}}{c_{y}}
\end{equation}
Regarding the puma, we have a qualitatively similar result. Again we have a ratio between their extinction rate ($e_y$) and reproduction ($c_y$). We also have an explicit ratio between the prey's extinction ($e_2$) and reproduction ($c_2$) rates, where, as expected, an increase in the prey's extinction rate decreases the final predator population, as opposed to an increase in the prey's reproduction rate, which increases the final puma population. We also have a simpler relationship between $y^*$ and $H$ now, we have a positive effect of the fraction not destroyed ($H$). 

Concerning sheep, we have a peculiar situation. Due to the initial mathematical formulation inspired by an automata model, we have a counterintuitive result. From tables \ref{tabela} and \ref{tabela2} we see that $e=e_y/a_2=e_y/(c_2H-e_2)$, so increasing the value of $e$ increases the pressure on the predator, which allows a larger population of prey to survive. But counterintuitively, once the conditions for this equilibrium point to be reached are met, increasing what was originally defined as the sheep's reproduction rate ($c_2$) decreases the final population of the prey in equilibrium and increasing their extinction rate ($e_2$), the final population is increased. There is also a negative effect of $H$ on the final sheep population.

Having analyzed the points where at least one of the species does not participate in the dynamics, it remains to see the conditions to have an equilibrium with the three species, which is what is pursued in sustainable equilibrium. In this situation, the three variables must be different from zero, which leads us to:
\begin{equation}
    \begin{split}
        0 & =	a-\frac{x_{1}^{*}}{\kappa_{1}}-\mu y^* \\
        0 & =	1-\frac{x_{2}^{*}}{\kappa_{2}}-y^*-px_{1}^* \\
        0     & =	x_{1}^*+x_{2}^*-e.
    \end{split}
         \label{eq:equilibrio3}
\end{equation}
 Solving the system we obtain for $x_1^*$:
\begin{equation}
    \begin{split}
x_{1}^{*}		& =\kappa_{1}\left(\frac{a+\frac{e\mu}{\kappa_{2}}-\mu}{1+\mu\left(\frac{\kappa_{1}}{\kappa_{2}}-\kappa_{1}p\right)}\right)
    \end{split}
\end{equation}
Or if we remember that $p=1/\kappa_1+1/\kappa_2$:
\begin{equation}
   \begin{split}
x_{1}^{*}		& =\kappa_1\left(\frac{a+\frac{\mu e}{\kappa_2}-\mu}{1-\mu}\right)
    \end{split}
    \label{eq6}
\end{equation}
Returning to the original parameters
\begin{equation}  \begin{split}         x_{1}^{*} & =\frac{a_{y}}{c_{1}}\left(\frac{\frac{a_{1}}{a_{2}}+\frac{\mu_{1}}{\mu_{2}}\frac{e_{y}}{a_{2}}\frac{c_{2}}{a_{y}}-\frac{\mu_{1}}{\mu_{2}}}{1-\frac{\mu_{1}}{\mu_{2}}}\right) \\& =\frac{c_{y}}{c_{1}}\left(\frac{\frac{c_{1}H-e_{1}}{c_{2}H-e_{2}}+\frac{\mu_{1}}{\mu_{2}}\frac{e_{y}}{c_{2}H-e_{2}}\frac{c_{2}}{c_{y}}-\frac{\mu_{1}}{\mu_{2}}}{1-\frac{\mu_{1}}{\mu_{2}}}\right) \\    \end{split}\end{equation}
And the other variables can be written in terms of $x_1^*$ as $x_{2}^{*}  =e-x_{1}^{*}$ and $y^{*} =\frac{a}{\mu}-\frac{x_{1}^{*}}{\mu \kappa_{1}}$. So the last equilibrium point is:
\begin{equation}
    \boldsymbol{x_{7}^{*}}=\left(x_{1}^{*},e-x_{1}^{*},\frac{1}{\mu\kappa_{1}}\left(\kappa_{1}a-x_{1}^{*}\right)\right)
\end{equation}
or returning to the original parameters:
\begin{equation}\begin{split}    \boldsymbol{x_{7}^{*}}	&=\left(x_{1}^{*},e-x_{1}^{*},\frac{1}{\mu\kappa_{1}}\left(\kappa_{1}a-x_{1}^{*}\right)\right) \\	&=\left(x_{1}^{*},\frac{e_{y}}{a_{2}}-x_{1}^{*},\frac{\mu_{2}}{\mu_{1}}\frac{c_{1}}{a_{y}}\left(\frac{a_{y}}{c_{1}}\frac{a_{1}}{a_{2}}-x_{1}^{*}\right)\right) \\	& =\left(x_{1}^{*},\frac{e_{y}}{c_{2}H-e_{2}}-x_{1}^{*},\frac{\mu_{2}}{\mu_{1}}\frac{c_{1}}{c_{y}}\left(\frac{c_{y}}{c_{1}}\frac{c_{1}H-e_{1}}{c_{2}H-e_{2}}-x_{1}^{*}\right)\right)\end{split}\end{equation}
We can see that this last equilibrium point has a more difficult behavior to analyze by looking at its parameters and returning to the original parameters does not simplify the issue. The parameters have a complex relationship with each other, making it more challenging to predict the influence of each parameter on the final population of the species at equilibrium.  Only a few trivial behaviors can be directly analyzed. We can see that an increase in the final guanaco population $x_1$ has a negative effect on the final sheep population $x_2$. This same association, albeit to a different degree, also appears in the final population of predators. $y$ A possible first explanation is to look at it the other way round: it is not that the increase in the guanaco population $x_1$ is the cause of the decrease in the puma population $y$,  but rather the consequence.   A more detailed analysis of the effect of each parameter on the survival of each species will be possible when we use the proposed analysis method with the perceptron, as discussed in Section \ref{influence}. 

With the seven equilibrium points, we can linearize the system around each one and calculate the eigenvalues to determine which equilibria are stable. The result of exploring the system's five independent parameters will be displayed in n-ary graph. A ternary graph is a well-known triangular diagram in which each point corresponds to a specific combination of the ratio of three variables to a constant sum.  It is a way to visualize 3 variables in only 2 dimensions. 

\begin{figure}[ht]
\centering
\includegraphics[width=0.5\textwidth]{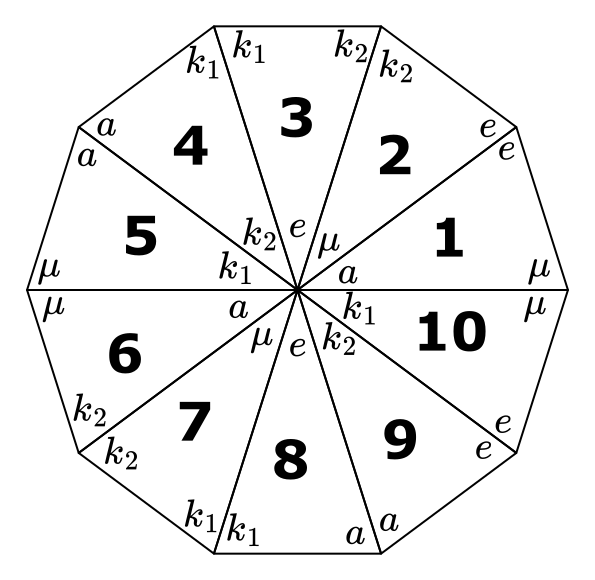}
\caption{The n-ary diagram is composed of multiple ternaries rearranged in a circular shape, the graph illustrates how the parameters are distributed in each of the ternary graphs. \label{pizza}}
\end{figure}

The n-ary graph consists of a rearrangement of multiple ternary graphs into a circular arrangement as shown in Fig. \ref{pizza}. Since in each ternary diagram three parameters vary and two are fixed, we have ten possible ternaries among the five parameters.
Since there are always two fixed parameters in each ternary, we define a set of ``default'' parameters that correspond to the values that the parameters will remain constant throughout the paper unless explicitly stated otherwise.

We define a set of parameters that result in an interesting state, i.e. in which the three species coexist, with a higher value for sheep and a lower for puma. That is an equilibrium state where $y<x_1<x_2$.  This is a result consistent with the real situation in Argentine Patagonia since sheep farming activity is predominant.  This status is obtained through the parameters  $a=0.40$, $\mu=0.6$, $\kappa_1=\kappa_2=1.43$, $e=0.7$,  The numerical solution was obtained using the second-order Runge-Kutta method with $\Delta t=0.01$, the initial populations of all species were set to $0.3$.
\begin{figure}[ht]
\centering
\includegraphics[width=0.7\textwidth]{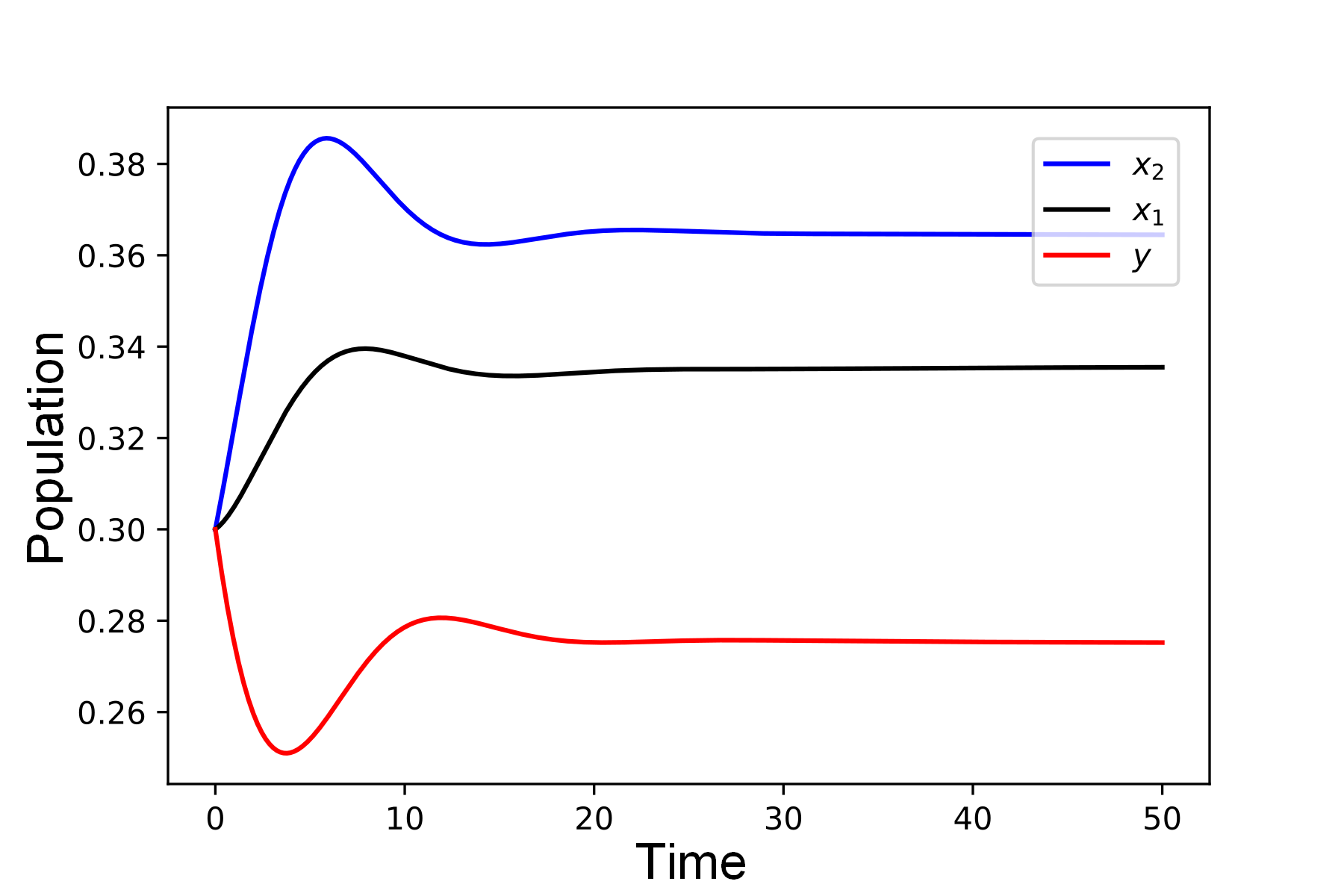}
\caption{Temporal evolution of the system of three species: guanaco ($x_1$), sheep ($x_2$) and puma ($y$). \label{evol}}
\end{figure}

A model inspired by the local ecosystem must incorporate some relationships based on real observations of this ecosystem, so we define that the predation probability of the sheep is higher than the predation probability of the guanaco, due to the hunting preferences of the puma.  Due to the hierarchical competition, it is necessary to give some ``advantage'' to the inferior competitor to allow its survival. This is usually done by giving the inferior competitor a higher reproductive rate than the superior competitor\cite{laguna1}

As discussed in \ref{MBE}, we have that $a=\frac{a_1}{a_2}$ is the ratio between the reproduction rates of the superior and inferior competitors. Similarly, $\mu=\frac{\mu_1}{\mu_2}$ is the ratio of the predation rate of the superior competitor to the inferior competitor. Therefore, since $a_2>a_1$ we obtain $a<1$. In an analogous way, we have that $\mu_2>\mu_1$ resulting in $\mu<1$. 

The parameter $e=\frac{e_y}{a_2}$ is the ratio between the extinction rate of the predator and the reproduction of the inferior competitor. In nature, it is expected to observe $e_y < a_2$ and consequently $e<1$, since new prey are born at a higher rate than predators die. 

However, the puma is a keystone species. A keystone species is a concept that characterizes the existence of species that, despite their low abundance in an ecosystem, play a critical role in ecological dynamics. The removal of a key species causes a drastic change in the ecosystem, in most cases, this species is a predator that controls the distribution and population of a large number of species that serve as prey. There are experiments to manage ecosystems based on a top-down control focused on key species, so this is an interesting parameter to extrapolate to incorporate the effects of increased pressure on puma survival due to human hunting~\cite{laguna1}.

The last parameters $\kappa_j$ are the carrying capacity related to each prey. These parameters do not have a justification to delimit a range of values, since they do not depend on the relationship between two distinct species. They are related to the capacity of the ecosystem to maintain a given population of a given species, i.e. the higher the value of the parameter the higher the population that the ecosystem can sustain of the respective species.   Mathematically this is evidenced when we analyze the equilibrium points, and looking at the equilibrium points for the case where only the prey survives. We have the sheep as an example that its final state is given by $\boldsymbol{x_2^*}=\kappa_2$. Thus, this is a value associated with the relationship of the species with the natural resources in particular of the ecosystem being modeled, rather than with the relationship with another species.

The carrying capacity values are usually greater than 1, to keep a standardization of the parameters (positive, less than 1 except $e$ and in the numerator) we will define a new parameter  $k_j = 1/\kappa_j <1$. From an ecological point of view, we are closer to Levins' models, a set of models mathematically equivalent to the logistic model that chooses to rewrite the logistic term so that all parameters are in the numerator. 

\begin{figure}[ht]
\centering
\includegraphics[width=0.6\textwidth]{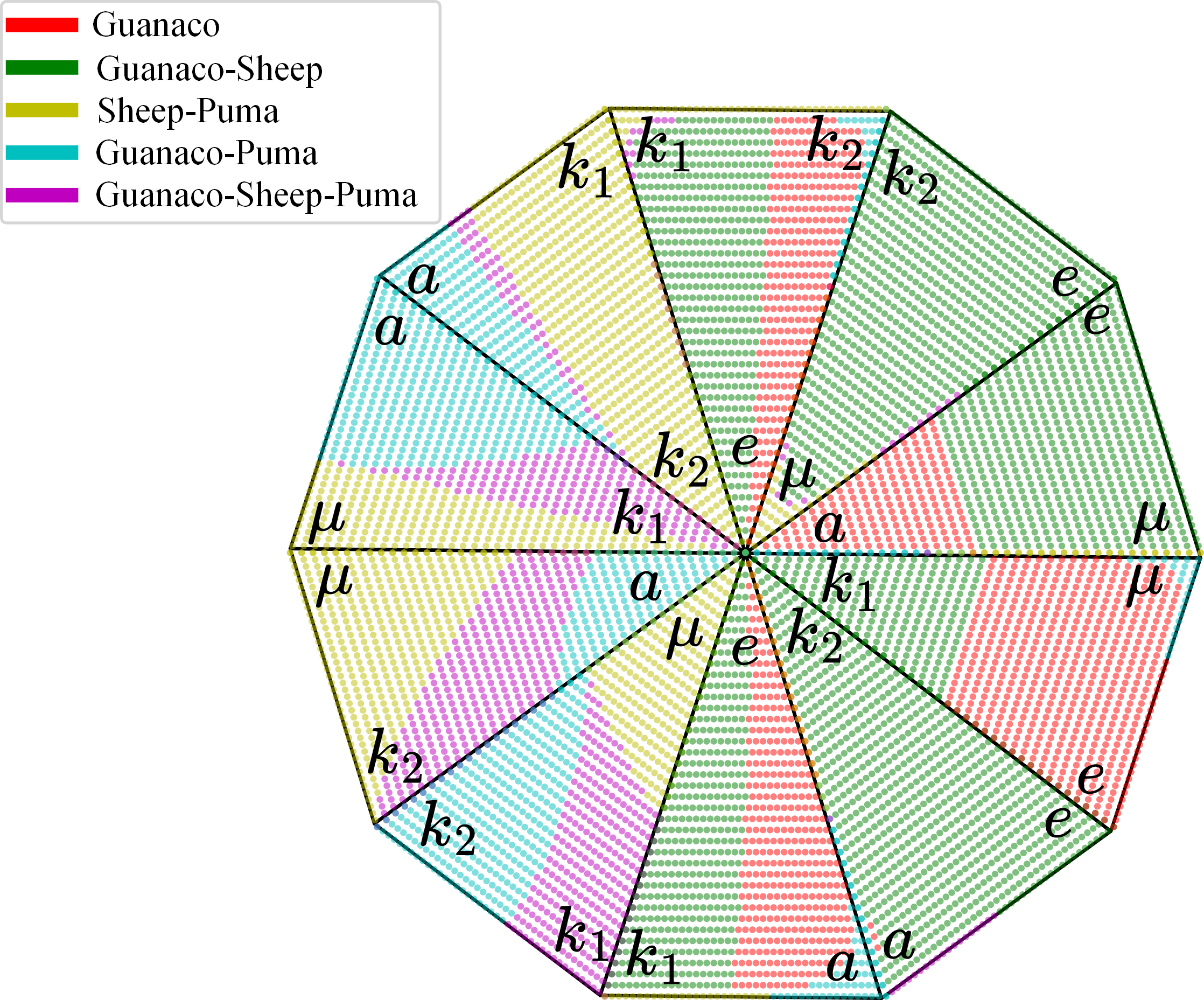}
\caption{N-ary diagram of the equilibrium states of the system in the defined parameter space. \label{fig2}}
\end{figure}

To represent the system through an n-ary graph, a color system was used to distinguish the different final states of the system. In Fig. \ref{fig2}, the normalized parameters vary between $1/100$ and $99/100$, while $e$ varies between $1/100$ and $100$.  Having defined the standard state and the range of values, we proceed by calculating the equilibrium points of the system, linearizing the system around each of these points, and calculating their respective eigenvalues. If the real part of all eigenvalues is negative, it turns out to be a stable equilibrium. 

In ternary 1 of Fig. \ref{fig2}, drawing a line parallel to the outer edge, the parameter $a$ remains constant since its value depends on the distance to the center of the polygon. For the states on the line where $a=0.5$, all equilibrium points have at least one eigenvalue with zero or positive real part. This happens on the boundary between the two phases  \footnote{Region in the parameter space in which the system evolves to the same equilibrium point}  of the system, Guanaco-Sheep, and Guanaco. In these parameter sets, numerical solutions show that the system always evolves to an equilibrium where the eigenvalues have a negative real part or zero. This equilibrium point corresponds to one of the phases that make the boundary in this region, specifically the isolated survival of the guanaco, that is, it does not correspond to a third phase on the boundary. No case with more than one stable equilibrium point was found.

We can also notice in Fig.~\ref{fig2} that the system exhibits five of the seven possible states, two states are not observed. These missing states are the extinction of the three species and the state with only sheep. The first case is an exceptional situation that we can achieve when the predator extinction rate is very low. 


 This result may seem counterintuitive at first. What happens is that the extinction rate of the predator is so low that in a short time, there is practically no extinction of the predator. This leads to an overpopulation of predators that drives all prey to extinction. So even if the extinction rate is very low, the predator population declines to extinction because there is no more prey, resulting in the collapse of the entire ecosystem. This is an unstable saddle-type equilibrium point, while the system approaches the equilibrium point if it is exactly on the axis that corresponds to the existence of puma alone, for any other situation than this,  that is, any disturbance caused by the introduction of some prey, the system then moves away from the equilibrium point.

The absence of the second end state, on the other hand, demonstrates the difficulty for the sheep to remain as the only surviving species. This is a consistent result since the animal is not native to the region, which results in it not only being an easier prey to the predator but also an inferior competitor to other native herbivores.

\begin{figure}[ht]
\centering
\includegraphics[width=0.7\textwidth]{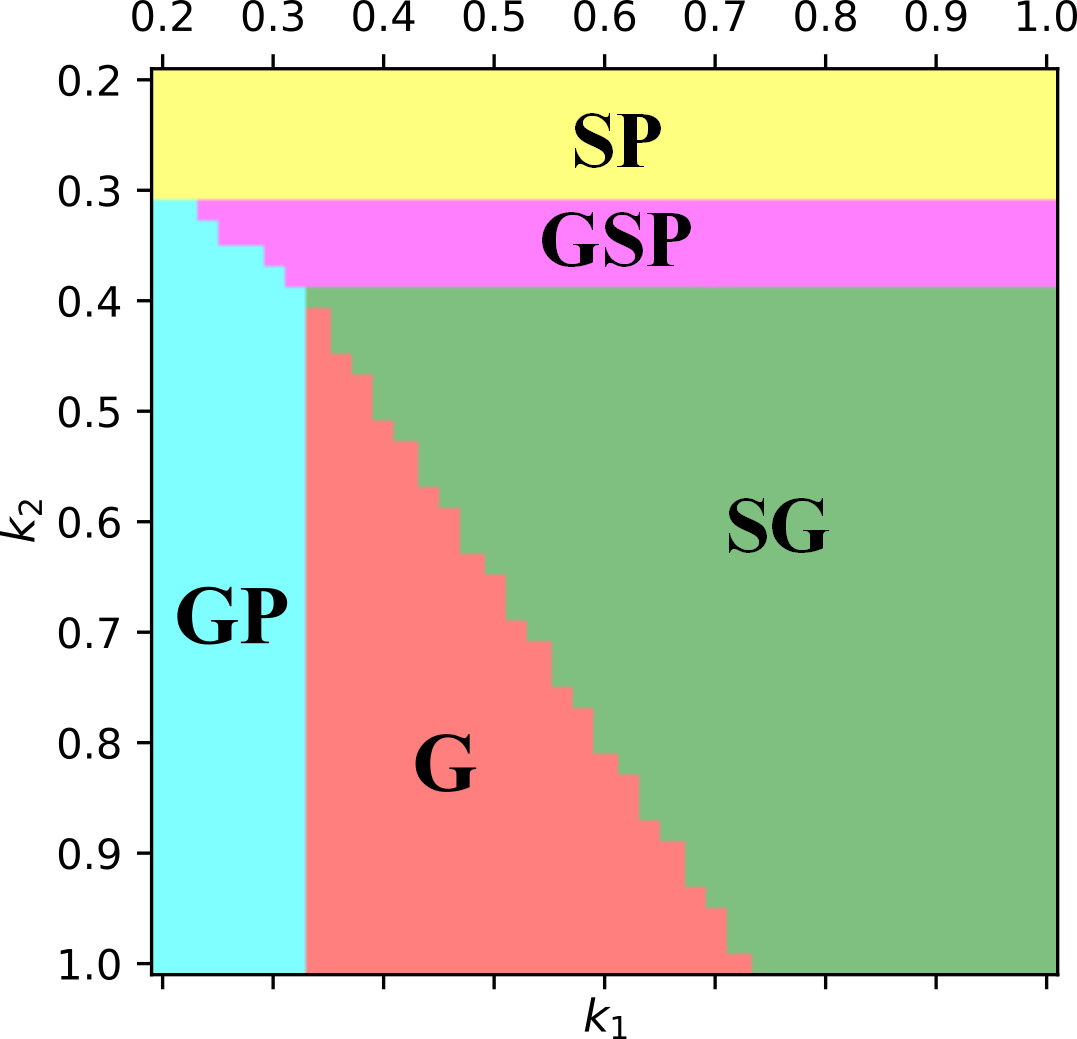}
\caption{Phase space where each phase is identified by the initials of the animals that coexist in the equilibrium state.\label{fig6}}
\end{figure}

It is also interesting to notice that in the third ternary graph in Fig. \ref{fig2} in a single line near the outer edge, it is possible to find all the $5$ states that are obtained in the graph. From a set of parameters, varying only $k_1$ and $k_2$ between $0$ and $1$ we then obtain the result that can be visualized in Fig. \ref{fig6} in which the system displays the 5 possible states.

To generate this Fig. \ref{fig6}, a process analogous to Fig. \ref{pizza} was employed. We can notice that all states can still be found even if we limit $k_1$ and $k_2$ between $0$ and $0.5$, this region is where the system presents a greater diversity of states. If we observe only $0.5 \le k_j \le 1$ we observe only two states, or even if $k_j >0.8$ there is then a single state to be observed.  The parameters $k_j$ seem to have more influence on the survival of each species at low values. For higher values these parameters alter the behavior of the system, specifically altering its oscillatory behavior, but no longer interfering with the survival or otherwise of each species, which is the focus of our analysis.

\subsection{Influence of each parameter}
\label{influence}
The interpretation of the system parameters in the survival of each species obtained only via intuition is limited and qualitative as we realised in the section \ref{equiibrio}. We expect that increasing $a$ has a positive effect on the survival of $x_1$ and that $e$ has a negative effect on the survival of $y$, but we may have difficulty inferring via intuition whether $a$ affects $x_1$ more than $e$ affects $y$, or how each of the parameters affects other species. This requires us to develop a quantitative, albeit approximate, measure of the importance of each parameter.

A greater understanding of the parameters leads to a better understanding of the model. Artificial neural networks are some of the most popular machine learning algorithms today. These algorithms have been used in many different areas largely due to their enormous success in different tasks, among them, binary classification. Even the simplest of models, such as the perceptron together with the learning algorithm, has evidence of being powerful enough for the model to converge to a solution in a finite number of steps since the solution exists~\cite{hagan}. 

\begin{figure}[ht]
\centering
\includegraphics[width=0.9\textwidth]{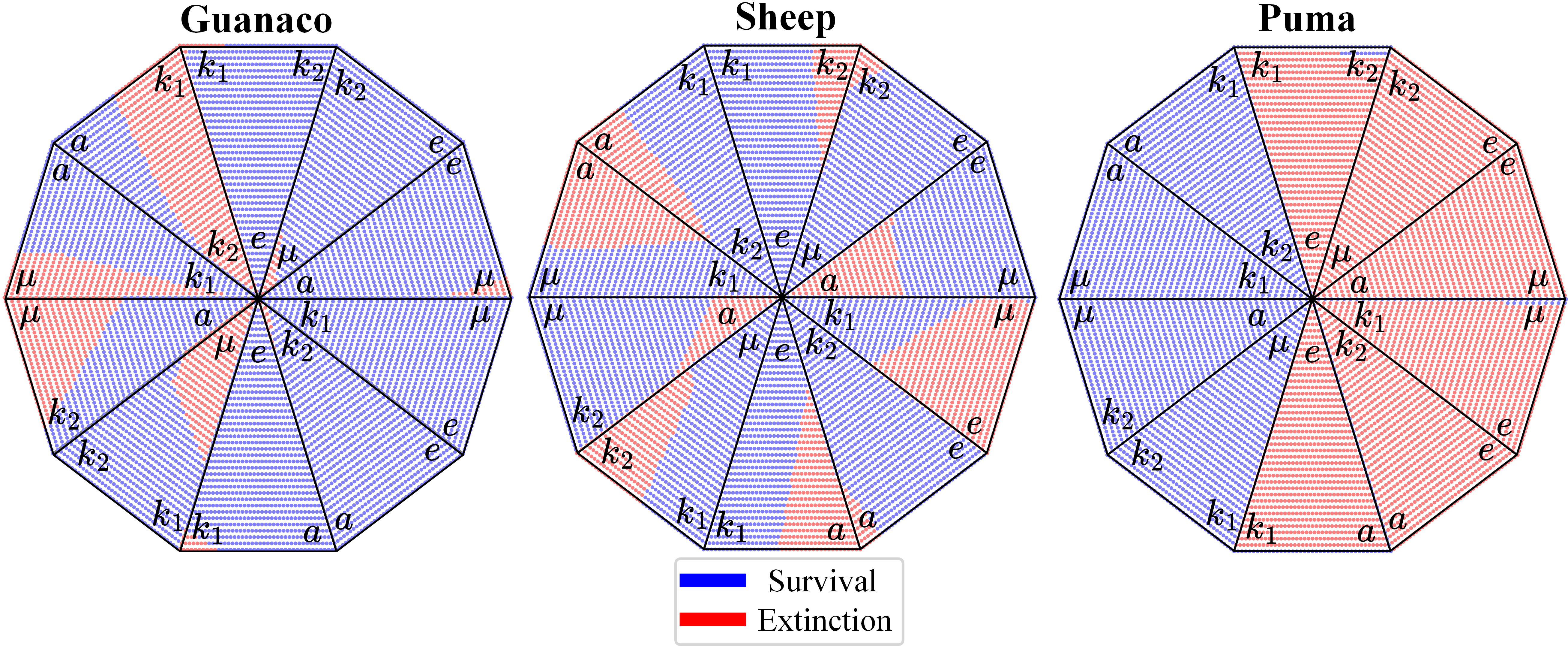}
\caption{Survival graph for each individual species.  \label{fig8} }
\end{figure}

Figure \ref{fig8} illustrates the individual survival of each species. We can notice that there is an almost linear separation between the two possible outputs. This result encourages us to hypothesize that a linear classifier like the perceptron can provide us with a classification with good accuracy. We then use a perceptron for each species, always providing as input the set of parameters $\left\{ a,\mu,e,k_{1},k_{2}\right\}$ and obtaining as output the classification of the species between two classes $\left\{ 0, 1\right\}$, alive or extinct.

Our main interest does not lie in the prediction ability of the perceptron. Machine learning models are essentially mathematical models, where more complex models can become abstract and difficult to interpret, but the simplicity of the perceptron can be seen as an advantage in terms of transparency.

The perceptron aims to predict the survival status of the species in the proposed model by taking the system parameters as inputs, thus the weights are a quantitative measure of the importance that the parameter associated with the weight has for the survival of the respective species of the perceptron. A negative weight indicates a negative contribution to the survival of the species, that is, a contribution to its extinction. An accessible understanding of the perceptron can be found in the specific literature on the topic\cite{hagan,nielsen}.

It should also be noted that more important than the magnitude of each weight are the relative values between them (and the bias). Thus, for better comparison between the different perceptrons, after the training process, each neuron had its bias normalized and the weights adjusted proportionally.

\begin{figure}[ht]
\centering
\includegraphics[width=0.7\textwidth]{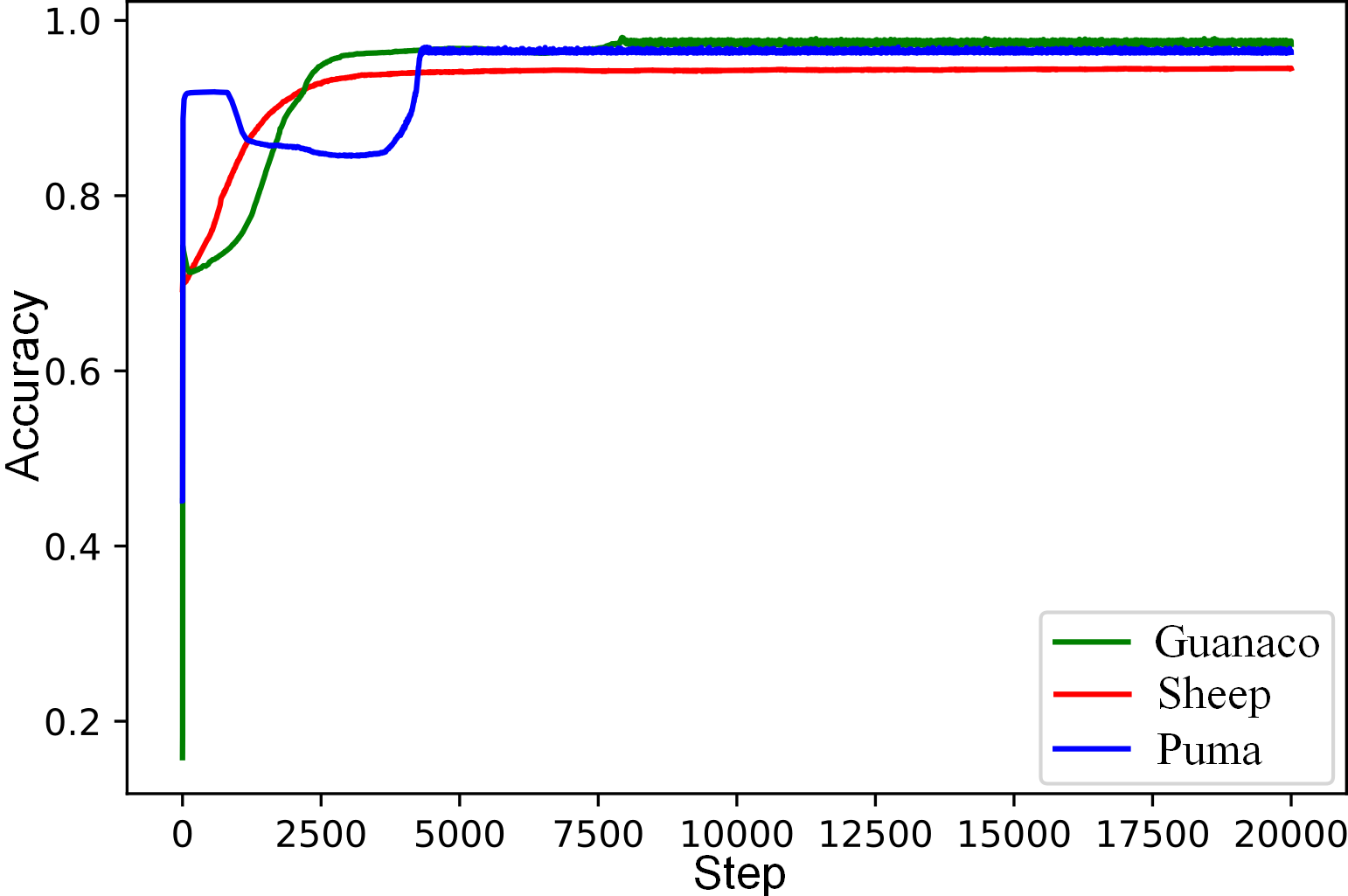}
\caption{Evolution of accuracy throughout the perceptron training stage. \label{treinamento}}
\end{figure}

The training dataset was generated with a process analogous to the one used in Fig. \ref{fig2}, saving the set of parameters of each point and the corresponding final state. By dividing each range of values into 40 points, we constructed a training dataset with $8610$ points. The training was performed with a learning rate of $\eta= 10^{-6}$ and running the entire training set $20000$ times.  The final classification accuracy was above 90\% for the three animals as we can see in Fig. \ref{treinamento}.

Accuracy was computed as follows: With each new input the perceptron generates an output and then registers whether the predicted output was correct or not, and if not, updates the weight. Always after testing the entire training dataset, we recorded the percentage of correct predictions for this run. Between one input and another, the weight is updated and the accuracy is computed only after we run through the whole training set. Consequently, the weight was updated several times during this process and the accuracy computed at each run does not necessarily correspond to the accuracy that would be computed if we tested the training dataset adopting a fixed weight vector obtained at the end. This difference is as large as $\eta$ due to the variation that the weight vector undergoes with each new wrong entry. 

The training result is summarized in table \ref{tabela5} as a normalized confusion matrix. This result was obtained by testing the entire training dataset keeping the weight vector constant, so this is a more accurate measure of the final accuracy. Each position in the cell corresponds to one of the 4 possible cases, and the main diagonal corresponds to the correct classifications. The 4 possible situations are:
\begin{itemize}
    \item The ideal output was $0$ and the output obtained was $0$: correct classification;
    \item The ideal output was $1$ and the obtained output was $1$: correct classification;
    \item The ideal output was $0$ and the obtained output was $1$: incorrect classification;
    \item The ideal output was $1$ and the output obtained was $0$: incorrect classification.
\end{itemize}

\begin{table}
\centering
\begin{tabular}{cccccccccccc}
                                              & \multicolumn{3}{c}{Guanaco}                                                    &                       & \multicolumn{3}{c}{Sheep}                                                      &                       &                        & Puma                      &                           \\ \cline{3-4} \cline{7-8} \cline{11-12} 
                                              & \multicolumn{1}{c|}{}  & \multicolumn{2}{c|}{Ideal}                            &                       & \multicolumn{1}{c|}{}  & \multicolumn{2}{c|}{Ideal}                            &                       & \multicolumn{1}{c|}{}  & \multicolumn{2}{c|}{Ideal}                            \\ \cline{3-4} \cline{7-8} \cline{11-12} 
                                              & \multicolumn{1}{c|}{}  & \multicolumn{1}{c|}{0}    & \multicolumn{1}{c|}{1}    &                       & \multicolumn{1}{c|}{}  & \multicolumn{1}{c|}{0}    & \multicolumn{1}{c|}{1}    &                       & \multicolumn{1}{c|}{}  & \multicolumn{1}{c|}{0}    & \multicolumn{1}{c|}{1}    \\ \cline{1-4} \cline{6-8} \cline{10-12} 
\multicolumn{1}{|c|}{\multirow{2}{*}{Output}} & \multicolumn{1}{c|}{0} & \multicolumn{1}{c|}{0.18} & \multicolumn{1}{c|}{0.02} & \multicolumn{1}{c|}{} & \multicolumn{1}{c|}{0} & \multicolumn{1}{c|}{0.25} & \multicolumn{1}{c|}{0.06} & \multicolumn{1}{c|}{} & \multicolumn{1}{c|}{0} & \multicolumn{1}{c|}{0.55} & \multicolumn{1}{c|}{0.00} \\ \cline{2-4} \cline{6-8} \cline{10-12} 
\multicolumn{1}{|c|}{}                        & \multicolumn{1}{c|}{1} & \multicolumn{1}{c|}{0.01} & \multicolumn{1}{c|}{0.79} & \multicolumn{1}{c|}{} & \multicolumn{1}{c|}{1} & \multicolumn{1}{c|}{0.02} & \multicolumn{1}{c|}{0.67} & \multicolumn{1}{c|}{} & \multicolumn{1}{c|}{1} & \multicolumn{1}{c|}{0.02} & \multicolumn{1}{c|}{0.43} \\ \cline{1-4} \cline{6-8} \cline{10-12} 
\end{tabular}
\caption{Normalized training confusion matrix. \label{tabela5}}
\end{table}

To validate the training we built a validation dataset with $4960$ entries. This dataset was generated analogously to the training data, but in order not to repeat the same points, we changed the parameters between $2/100$ and $98/100$, except for $e$ which varied between $2/100$ and $99/100$ and the range of values was divided into $30$ points. The result is summarized in the table \ref{tabela7} and within the decimal places displayed the same result obtained for the training set.

\begin{table}
\centering
\begin{tabular}{cccccccccccc}
                                              & \multicolumn{3}{c}{Guanaco}                                                    &                       & \multicolumn{3}{c}{Sheep}                                                      &                       &                        & Puma                      &                           \\ \cline{3-4} \cline{7-8} \cline{11-12} 
                                              & \multicolumn{1}{c|}{}  & \multicolumn{2}{c|}{Ideal}                            &                       & \multicolumn{1}{c|}{}  & \multicolumn{2}{c|}{Ideal}                            &                       & \multicolumn{1}{c|}{}  & \multicolumn{2}{c|}{Ideal}                            \\ \cline{3-4} \cline{7-8} \cline{11-12} 
                                              & \multicolumn{1}{c|}{}  & \multicolumn{1}{c|}{0}    & \multicolumn{1}{c|}{1}    &                       & \multicolumn{1}{c|}{}  & \multicolumn{1}{c|}{0}    & \multicolumn{1}{c|}{1}    &                       & \multicolumn{1}{c|}{}  & \multicolumn{1}{c|}{0}    & \multicolumn{1}{c|}{1}    \\ \cline{1-4} \cline{6-8} \cline{10-12} 
\multicolumn{1}{|c|}{\multirow{2}{*}{Output}} & \multicolumn{1}{c|}{0} & \multicolumn{1}{c|}{0.18} & \multicolumn{1}{c|}{0.02} & \multicolumn{1}{c|}{} & \multicolumn{1}{c|}{0} & \multicolumn{1}{c|}{0.25} & \multicolumn{1}{c|}{0.06} & \multicolumn{1}{c|}{} & \multicolumn{1}{c|}{0} & \multicolumn{1}{c|}{0.55} & \multicolumn{1}{c|}{0.00} \\ \cline{2-4} \cline{6-8} \cline{10-12} 
\multicolumn{1}{|c|}{}                        & \multicolumn{1}{c|}{1} & \multicolumn{1}{c|}{0.01} & \multicolumn{1}{c|}{0.79} & \multicolumn{1}{c|}{} & \multicolumn{1}{c|}{1} & \multicolumn{1}{c|}{0.02} & \multicolumn{1}{c|}{0.67} & \multicolumn{1}{c|}{} & \multicolumn{1}{c|}{1} & \multicolumn{1}{c|}{0.02} & \multicolumn{1}{c|}{0.43} \\ \cline{1-4} \cline{6-8} \cline{10-12} 
\end{tabular}
\caption{Normalized validation confusion matrix. \label{tabela7}}
\end{table}

\begin{figure}[ht]
\centering
\includegraphics[width=0.9\textwidth]{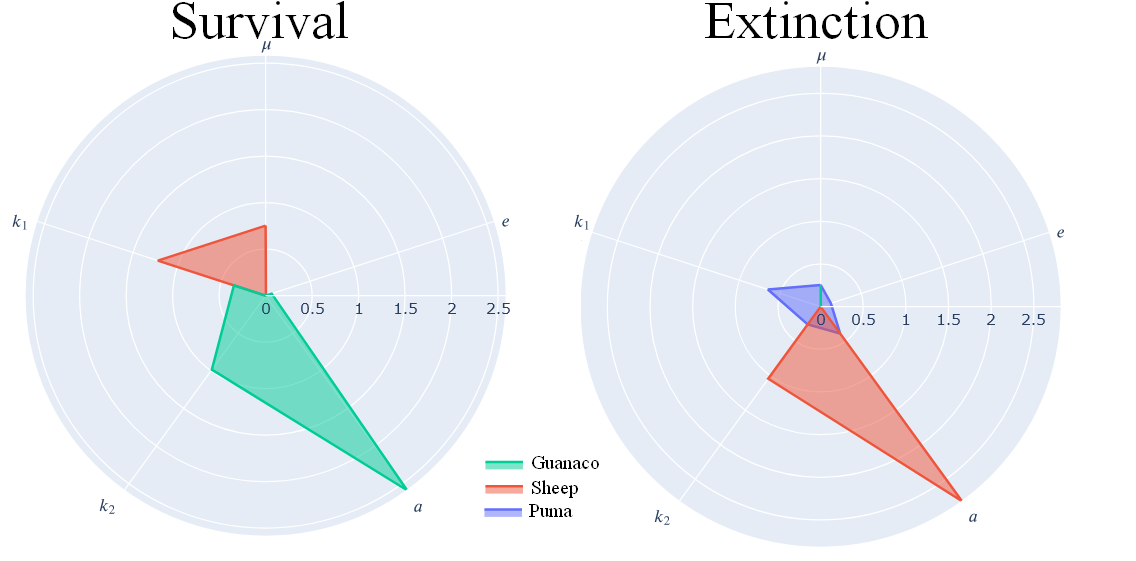}
\caption{Weights associated with each parameter are arranged in a radar chart separated between the effect on survival and extinction of each species.\label{teia}}
\end{figure}

The normalized weight vectors obtained at the end of the training, where the last element is the bias, are:

\begin{equation}
    \begin{split}
       \overrightarrow{w}_{x_{1}}& =\left(2.59,-0.97,0.07,0.35,0.98,-1\right) \\
\overrightarrow{w}_{x_{2}}& =\left(-2.82,0.75,0.00,1.21,-1.04,1\right) \\
\overrightarrow{w}_{y}& =-\left(0.39,0.25,0.13,0.65,0.26,-1\right)
    \end{split}
    \label{simplificado}
\end{equation}
The weights can be better visualized in Fig. \ref{teia} where we separate the visualization of each parameter for the survival or extinction of the species.

The mathematical equation describing the three perceptrons can be seen in equation \ref{per}, where there is survival of the species in cases where the final sum result is greater than 0, i.e., the guanaco survives if $x_1>0$.

\begin{equation}
\begin{split}
x_{1} & =2.59a-0.97\mu+0.07e+0.35k_{1}+0.98k_{2}-1  \\
x_{2} & =-2.82a+0.75\mu+1.21k_{1}-1.04k_{2}+1 \\
y  &  =-0.39a-0.25\mu-0.13e-0.65k_{1}-0.26k_{2}+1. \\
\label{per}
\end{split}
\end{equation}


It can be seen that the weight of $e$ for all the perceptrons is the lowest of all the parameters, reaching zero for $x_2$, so we can conclude that this is the parameter to which the system is least sensitive. Tests carried out for values between $0$ and $1$ for $e$ indicate the insensitivity of the system, showing no state in which the puma is extinct, for example.

For a positive bias, for the species to be extinct, the sum of the products $z_j\cdot w_j$, where $z_j$ is any parameter and $w_j$ is the associated weight, must be less than $-1$, the inverse for a negative bias.  In the scenery analyzed, the product $z_j\cdot w_j$ ( with the exception of $z_j=e$.), will have as a maximum result the value of the weight $w_j$ itself, since the parameter $z_j$ is always less than one, this helps us to better interpret the importance of each weight regardless of the parameter values.

If we look at the weight of parameter $a$ it is greater than $2$ for sheep and guanaco besides having the opposite sign of the bias, which implies that it has a great influence on their survival. That is, if we vary only this parameter between the extreme values ($0\leq a \leq1$) we expect to change the survival status of both species.  It is also interesting to note that although $a$ is the parameter with the highest magnitude for both prey, its influence works in opposite ways between species. It indicates the existence of competition between the species and resistance to coexistence, since if we increase its value to facilitate the survival of one species, we make it more difficult for the other.  

The puma is the only animal in which all parameters are negative, but no single parameter has a magnitude greater than the bias itself, so it can be considered the least sensitive species to parameter variation since we need more extreme parameter values to cause their extinction. It is remarkable that the parameter $e$, in a counter-intuitive way, exerts less influence on the survival of the species than other parameters such as $a$ itself. 

\begin{figure}[ht]
\centering
\includegraphics[width=0.9\textwidth]{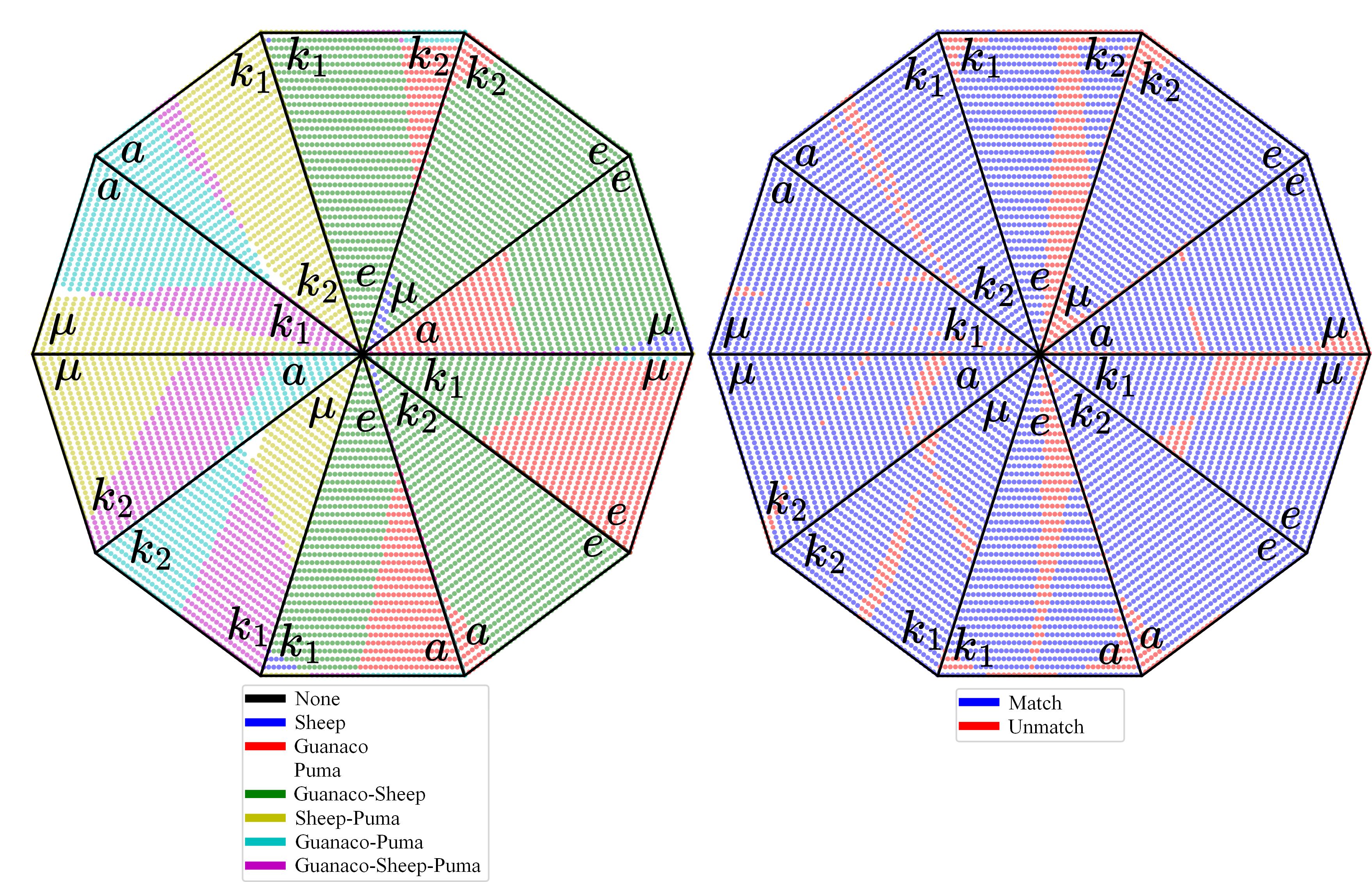}
\caption{On the left the result obtained via perceptron and on the right the comparison with the exact solution of the equation-based model.\label{comparacao}}
\end{figure}

Artificial neural networks of which the perceptron is part can be interpreted as an approximator of functions~\cite{hagan} . Using the linear system \ref{per} as a function approximator to identify the survival of each species as a function of the parameters, we get the result seen in Fig. \ref{comparacao} .

We can see that 87\% of the final states obtained by the perceptron match the analytical result obtained by the differential equation system. It can be pointed out that it is more difficult to correctly classify the final state of the system as a whole than the final state of each species individually. This stems from the fact that the classification is correct for the system only when it is correct simultaneously for all three species so that the accuracy can only be equal to or lower than when compared to the highest accuracy obtained individually for each species. 
  
Both models were run on "Collaboratory" via the free plan which provides a Google Compute Engine back-end in Python 3 with 12.7GB of RAM available, this is an online service from Google that allows you to write and run Python in the browser. While Fig. \ref{fig2} required 21 minutes and 1 second to generate, Fig. \ref{comparacao}, required only 2s.

\section{Discussion}


There are two regions in the graph \ref{fig2} where a scenario already known in literature occurs: starting from a state of coexistence between sheep and puma, we then increase the extinction rate of the predator. As the predator population decreases, there is first an intermediate state where all animals coexist, and then for the most extreme values of $e$ the system reaches a final state where the puma population is extinct and only the two prey remain~\cite{laguna1}.

This transition appears both in ternary $3$ and $2$. The ternary $2$ draws particular attention because the transition occurs on an edge of the triangle, i.e., the third parameter remains constant, thus only $e$ and $\mu$ vary.  Analyzing this edge one can notice that as we move away from the center of the graph, the value of $e$ increases and we go through the sequence described above.

This result can also be interpreted by looking at the weights obtained via perceptron. The parameter $e$ has its lowest influence on the sheep, a species that does not change its status throughout the transition. This parameter has a positive influence on the guanaco and a negative influence on the puma, so as the value of the $e$ parameter increases, the puma population is driven to extinction at the same time as the guanaco ceases to be extinct, while the sheep remains unchanged.

Another well-known result from the literature already discussed is the total extinction of all species for $e\approx 0$ with $e \neq 0$~\cite{laguna1}. These results were obtained with a model based on differential equations with almost half the parameters of the original models.

Thus, by combining the analyses performed with the perceptron and the n-ary, it was possible to go beyond merely retrieving results found in the literature. Another good illustration of the consistency between the analyses can be obtained when observing the sixth ternary in Fig. \ref{fig2}. In the center of the circle, there is the maximum value of $a$, when only guanacos and pumas coexist, moving away from the center towards the edge, the value of $a$ suffers a decrease causing the transition of the system to a state where only sheep and pumas coexist, passing again through the state of coexistence between the three species. Analogous to the situation discussed previously. Looking at Fig. \ref{teia}, we notice exactly that $a$ has a strong positive influence on the survival of the guanaco and a negative one on the sheep, and finally is less felt by the puma. 

Through Fig. \ref{teia}, we can clearly see how each parameter acts on each species. This quantitative information is the main information that the perceptron brings us about the system, and through the radar chart, we can quickly and efficiently visualize how each parameter affects each species. For example, when we look at the sheep, we can see that while $\mu$ and $k_2$ contribute to the survival of the species, in a relatively similar amount, $k_1$ and $a$ contribute negatively, in addition, we can see $a$ is more important than the other parameters. We can still note that $e$ is the least relevant. 

Similar analyses can be performed for all species. Comparing the three species, it is clear that the parameters behave in an opposite way for the sheep and guanaco, signaling a kind of competition between both species. We can also see that the Puma is the less sensitive animal to change parameters. In this way we were able to use the perceptron in an unusual way quite interesting for the analysis of the system, more interesting than the prediction capacity, the simplicity and transparency it allowed us to obtain a direct interpretation of the weights of each perceptron, associating the weight with the importance of each parameter to each species, and the choice of the appropriate way of visualizing this data allows us to translate this information into a very simple graphical representation. This information is not usually obtained in the more traditional methods employed in the field.

It was also possible to find a set of parameters in which by varying only two parameters within an identified range of values one can find the 5 equilibrium states, not obtaining only the two special cases discussed previously.  Looking at figure \ref{fig6}, keeping fixed $k_2=0.9$ and varying $k_1$ between $0$ and $1$ we start from the coexistence between guanaco and pumas, as we increase $k_1$ should decrease the final population of guanacos in the equilibrium state, but the predator population goes to extinction before than the guanacos because the low amount of prey, taking the system to the transition where we have only the guanaco alive. By increasing $k_1$ further we increase the pressure on the guanacos so that it reaches such a state that without predator there is no longer enough pressure to keep the sheep extinct and the system undergoes the transition to the state where both prey coexist. 

Thus, we can see that $k_1$ in Figure \ref{teia} presents a counterintuitive result that also favors the survival of the guanaco, although it decreases the final population of the species in the equilibrium state. That is $k_1$ within the range of parameters analyzed, drives the predator population faster to extinction than the prey population, eliminating the external pressure on the guanaco. If we fix $k_1=0.9$ and vary $k_2$ we have a similar effect, where the decrease of the sheep since it is the only living prey, leads to the extinction of the predator quickly, but the extinction of the predator allows the survival of the guanaco which is a superior competitor and also exerts a pressure against the survival of the sheep, in this way $k_2$ does not contribute to the survival of the sheep, but to its extinction, since it does not eliminate external pressure and decreases its own population more and more.

These are interesting results and quite difficult to obtain intuitively, the fact that it is possible to explore the 5 parameters of the system at the same time and obtain a quantitative measure on each parameter provides valuable tools so that it is possible to find these regions of greatest interest in the state space of the phases without or with little prior knowledge of the system.

Despite the simplicity of the perceptron model of artificial neural networks, it was possible to obtain good accuracy. Simplicity in this situation can be seen as an advantage since it allows a better understanding of the classification process, not being a ''black box''. This is an essential feature for the development of human-centered technologies. 

It is also important to emphasize that the choice of parameters and the range of values they can vary is usually made in an apparently arbitrary way. However, in our model, it was possible to present arguments based on observations in nature to limit the range of values of 3 of the 5 parameters, and for the remaining 2 we presented a hypothesis as to why increasing the range of values would not bring any additional benefit to the analysis.

It should also be noted that according to the literature on the subject and the ecosystem in particular, it was considered that the $p$ parameter responsible for regulating hierarchical competition between prey is not an independent parameter, but is written in terms of other parameters present in the system. An obvious generalization is to treat this parameter independently. 

In this work, only a few particular cases were found in which it was not possible to obtain a single state in which all eigenvalues are negative, yet numerical solutions indicated the existence of a single equilibrium point. It is possible to find situations that deviate from this pattern with other value ranges or other models. These exceptions may identify interesting regions to explore. For example, finding regions where the equilibrium state of the system depends on the initial population since the system has more than one equilibrium point. This result was not observed in this model, but it is a possibility of investigation that is usually not explored, since the method traditionally adopted is only to maintain a fixed initial population and numerically solve the system for a range of parameter values, ignoring possible information about other equilibrium points.

Regarding the perceptron, we adopted the simplest possible linear classification model, partly for simplicity of application, and partly for simplicity of interpretation. However, it is natural to think about the possibility of experimenting with more complex artificial neural network models with greater classification capacity, including the possibility of performing non-linear classifications. In this way, we could build approximators of functions to describe the behavior of more sophisticated models.

For example, in Fig. \ref{fig6} we can notice a non-linear behavior specifically in the behavior of the puma, with a non-linear classifier we could have a better accuracy in this region of the parameter space, maybe even a non-linear classifier obtained from the combination of only two linear classifiers could present a better accuracy for the region. Keeping linear classifiers we could still address strategies to further improve the classification, for this we can mention the possibility of using other training rules and even different learning rates. Another option for future development is to keep the same parameters and take advantage of the fact that the results can be obtained faster through the trained perceptron when compared to the initial set of equations, and try to extrapolate the training to a larger parameter space.

To finish, it is worth having a final discussion about the choice to focus attention on which species survive when the system reaches the equilibrium state over how the system evolves. Let us first recall the fact that the analytical method of obtaining the stable equilibrium point is computationally faster to obtain than numerically solving the system. Moreover, we do not have a precise way to deduce at which instant the system will be in equilibrium. We can check manually for each set of parameters, choose a fixed $\tau$ where we assume the system will be in equilibrium for any set of parameters, or else develop an additional code to do this monitoring. In any case, obtaining directly the equilibrium points is a faster, more accurate, and more practical solution when possible.

The second point is the fact that no parameter has a direct association with any observable measured in nature. This allows us to perform different manipulations and rearrange the system in different ways, but it also interferes with having an intrinsic interpretation of the meaning of each variable or parameter. In the dimensionless model that was developed, the variables representing each species $\left(x_{1},x_{2},y\right)$ were rewritten in terms of the product between the parameters and parameters of the system in its dimensional form. Therefore there is no intrinsic interpretation of what the variable $y$ associated with the cougar means. We could think of different interpretations, such as the number of animals, meta-populations, biomass, etc. But regardless of the interpretation, it is consistent to assume that since $y$ is associated with the puma population, if $y=0$ then the species is extinct, consequently if $y>0$ then the species is alive. 

Thus, more relevant than how the system evolves, and even the magnitude of the variables in the equilibrium state of the system, it is more interesting to know only if $y>0$ or $y=0$, that is, whether the puma survives or becomes extinct, the same reasoning holds for the other animals. This approach also proved to be more suitable to approach binary classification artificial neural networks, because using perceptrons, we cannot recover the evolution of the system, but we were able to obtain a linear function that can tell us approximately whether a species survives or not given a set of parameters.

In general, beyond the results obtained for the proposed model, the proposed analysis method brings together characteristics found in different works that focus on analyzing different aspects of the models.  Previous works adopt different strategies, each with its own limitations. When more general analysis is adopted, such as the identification and analysis of equilibrium around equilibrium points, or even the existence and characteristics of the exact solution of the system, it is necessary to work with simpler models, usually with fewer parameters and fewer species involved. On the other hand, when analyzing more complex systems, one usually focuses on more specific analyses, on only a few parameters and using numerical solutions. What we also propose with our model is to synergistically use different tools so that we can work with an intermediate case in a more complete way.

About the model, proposing a system with fewer parameters than previous works developed on Argentine Patagonia, but with more parameters and species than the simplest mathematical ecological models in general, using both numerical solutions, linearization around the equilibrium point, and new tools such as n-ary graphs and artificial neural networks, it was possible to obtain a deeper exploration of the proposed model. A more complete analysis was obtained than is normally expected in systems of similar mathematical complexity, a similar scope to that obtained in simpler systems, but without losing important behaviors obtained with more complex systems or the ecological interpretation. Thus, the proposed methodology also adds to the results of this paper being not only an appropriate methodology for the investigation of this particular model but something that can be generalized and adapted to work with different models in general.

Thus, we believe that this work has been successful in achieving its objectives and has the potential to contribute to the development of the field and to the improvement of the decision-making process.

\section*{Author contributions}
Jhordan Silveira de Borba performed   conceptualization,  data curation,  formal analysis,   investigation , methodology,   software,   visualization and  Writing (original draft, review and editing). Sebastian Gonçalves performed  conceptualization,  methodology,  project administration,  supervision and  Writing (review and editing).






\section*{Funding sources}
We acknowledge financial support CNPq for partially supporting this work under grant \#314738/2021–5 and grant \#130391/2021-2.


\end{document}